\documentclass{article}

\usepackage{spconf,amsmath,graphicx}

\usepackage{amsmath}
\usepackage{graphicx}
\usepackage[dvipsnames]{xcolor}
\usepackage{amssymb,amsfonts}
\usepackage{caption}
\usepackage{subcaption}
\usepackage{algorithmic}
\usepackage{algorithm}
\usepackage{cite}
\usepackage[export]{adjustbox}
\usepackage[footnote]{acronym}
\usepackage{diagbox}
\usepackage{bm}

\usepackage{array}
\usepackage{multirow}
\usepackage{hhline}
\makeatletter
\newcommand{\thickhline}{%
	\noalign {\ifnum 0=`}\fi \hrule height 1.1pt
	\futurelet \reserved@a \@xhline
}
\newcolumntype{"}{@{\hskip\tabcolsep\vrule width 1.1pt\hskip\tabcolsep}}
\makeatother

\usepackage[hidelinks]{hyperref}

\usepackage{siunitx}

\input{new_commands.tex}

\usepackage[acronym, nopostdot, nonumberlist, nogroupskip, style=super]{glossaries}

\newacronym{rr}{RR}{ridge regression}
\newacronym{tsvd}{TSVD}{truncated singular value decomposition}
\newacronym{lv}{LV}{Loris-Verhoeven}
\newacronym{fp}{FP}{Fabry-Pérot}
\newacronym{snr}{SNR}{signal-to-noise ratio}
\newacronym{fts}{FTS}{Fourier-transform spectroscopy}
\newacronym{opl}{OPL}{optical path length}
\newacronym{opd}{OPD}{optical path difference}
\newacronym{svd}{SVD}{singular value decomposition}
\newacronym{dct}{DCT}{discrete cosine transform}
\newacronym{imspoc}{ImSPOC}{Imaging SPectrometer On Chip}
\newacronym{cc}{CC}{ColorChecker}
\newacronym{rmse}{RMSE}{root mean squared error}
\newacronym{ctv}{CTV}{collaborative total variation}
\newacronym{tv}{TV}{total variation}
\newacronym{pnp}{PnP}{Plug-and-Play}
\newacronym{irca}{IRCA}{interferometer response characterization algorithm}
\newacronym{hsi}{HSI}{hyperspectral image}
\newacronym{admm}{ADMM}{alternating direction method of multipliers}
\newacronym{wn}{WN}{wavenumber}
\newacronym{ssim}{SSIM}{structural similarity index measure}
\newacronym{rgb}{RGB}{red green blue}
\newacronym{bm3d}{BM3D}{block-matching and 3D filtering}

\title{Spectro-spatial hyperspectral image reconstruction from interferometric acquisitions}
\name{
	Daniele~Picone,
	Mohamad~Jouni,
	Mauro~Dalla~Mura
	\thanks{
        Mauro Dalla Mura is also with Institut Universitaire de France (IUF), 75005 Paris, France.
		This work is partly supported by grant ANR FuMultiSPOC (ANR-20-ASTR-0006),
		and partly by Région Auvergne-Rhône-Alpes grant ``Pack Ambition International 2021'' (21-007356-01FONC, 21-007356-02INV).
	}
}
\address{
	Univ. Grenoble Alpes, CNRS,\\
        Grenoble INP, GIPSA-lab,\\
        38000 Grenoble, France
}

\begin{document}
%
\maketitle
\begin{abstract}
In the last decade, novel hyperspectral cameras have been developed with particularly desirable characteristics of compactness and short acquisition time, retaining their potential to obtain spectral/spatial resolution competitive with respect to traditional cameras. However, a computational effort is required to recover an interpretable data cube.
In this work we focus our attention on imaging spectrometers based on interferometry, for which the raw acquisition is an image whose spectral component is expressed as an interferogram. Previous works have focused on the inversion of such acquisition on a pixel-by-pixel basis within a Bayesian framework, leaving behind critical information on the spatial structure of the image data cube.
In this work, we address this problem by integrating a spatial regularization for image reconstruction, showing that the combination of spectral and spatial regularizers leads to enhanced performances with respect to the pixelwise case. We compare our results with Plug-and-Play techniques, as its strategy to inject a set of denoisers from the literature can be implemented seamlessly with our physics-based formulation of the optimization problem.
\end{abstract}

\section{Introduction}
\label{sec:introduction}

Characterizing the \textit{spectrum} of a light source (i.e., measuring its intensity at each wavelength) is at the core of \textit{imaging spectroscopy}, and has deep implications in various fields, such as geology, gas detection, security, remote sensing, disaster prevention, and more~\cite{Eism12, Mano16}.
In recent times, both the scientific community and industrial venues have shown interest in \gls{hsi} spectrometers which operate on the principle of \textit{interferometry}, as they potentially allow for instruments with reduced cost and dimensions and for acquisitions with finer spectral resolution and improved \acrshort{snr}~\cite{Hech16, Hari10}.

Hyperspectral cameras based on interferometry measure the interference of superimposed coherent light waves which travel across different optical paths, known as the \glspl{opd}. At the detector level, a grayscale image is formed for each value of the \gls{opd}. The technique of capturing data across different \glspl{opd} is called \gls{fts}. Finally, a datacube is obtained where each pixel of the observed scene a vector of values in the domain of \gls{opd} known as the \textit{interferogram}.

However, in order to obtain the \gls{hsi}, two issues arise.
First, a data processing step is required to reconstruct the spectral response in the domain of the wavelengths, as the acquisitions are only available in a transformed domain.
Second, the spatial information should be taken into account in the reconstruction as a pixel-to-pixel spectral reconstruction \cite{jouni2023model} ignores spatial noise and pixel neighborhood information.
Given the nature of this computational imaging~\cite{Dona19} problem, a combination of spectral and spatial regularizers is required.

%
Machine learning-based approaches may provide a way to intrinsically regularize this formulation, but the availability of training data is limited in our scenario. As we deal with prototype instruments, 
it is complicated to build a large enough database of coupled acquisitions and references.
To partly address this issue, hybrid (model- and learning-based) approaches and basic \gls{pnp} approaches \cite{kamilov2023plug}, such as algorithm unrolling \cite{MongLE21:spm} and deep priors \cite{ZhanLZZVT21:tpami}, are currently gaining traction in the community. They rely heavily on a good understanding of the system, which we aim to address here as preliminary step and solid basis for the \textit{interpretability} of such systems.

In this work, we propose a \textit{spectro-spatial} model-based inversion for the \gls{hsi} reconstruction of interferometric image acquisitions, by exploiting the knowledge of the physics and the characteristics of interferometric systems and by showing the importance of having priors that are adapted to the data.
This is an extension to our previous work~\cite{jouni2023model} which was only pixel-to-pixel based.

In this paper, we extend the study to combine spatial regularizers with the spectral ones thanks to \gls{ctv} \cite{DuraMSC16:jis}, applied on the case where a potentially infinite amount of interfering waves can be superimposed. For example, this is capable to also deal with several instruments based on \gls{fp} interferometers~\cite{zucco-2014-fabry-perot, oiknine-2018-multi-apert, pisani-2009-compac-fourier, picone2023interferometer}; this is defined in our work as $\infty$-wave model, also known as Airy distribution \cite{Hech16}.
We propose to solve the inversion problem through a simple enough solver, notably the \gls{lv} algorithm \cite{LoriV11:ip}, with proximal operators that better characterize the \textit{a priori} knowledge through induced sparsity on a spectro-spatial cascaded Fourier domain transform and \gls{tv} operations.
We compare our results with pixel-to-pixel spectral reconstruction \cite{jouni2023model} and \gls{pnp} techniques, especially when the acquisitions are affected by noise.
Our contributions can be summed up as follows:
(i) we propose a model-based \gls{hsi} reconstruction method from interferograms based on spectro-spatial regularization,
(ii) we perform a comparison with image priors based on \gls{pnp} techniques within the context of imaging spectrometers based on interferometry, to show the importance of having priors that are adapted to the data.

\section{Proposed methodology}
\label{sec:inversion}

\subsection{Problem statement}

    
    We want to set up the reconstruction problem as a Bayesian problem. Let $\tens{X}\in\R{I \times J \times K}$ describe an unknown datacube to reconstruct representing the scene with $I \times J$ pixels and $K$ channels.
    The acquisition process is then described as a stochastic process in the form:
    \begin{equation}
    	\tens{Y} = \mathbb{A}(\tens{X}) + \tens{E}\;.
    	\label{eq:direct_model}
    \end{equation}
    In the above equation, $\tens{E}$ denotes additive white Gaussian noise and $\tens{Y}\in\R{I \times J \times L}$ describes our acquisition. The latter retains the same spatial resolution of our target datacube, while its spectral content is defined in a transformed domain with $L$ samples. This domain transformation is defined by a linear operator $\mathbb{A}$, which we define as:
        \begin{equation}
    	\mathbb{A}(\tens{X}) = \mathbf{A} \con\nolimits_3\tens{X}\;,
    	\label{eq:direct_domain_transform}
    \end{equation}
    where $\con_3$ denotes a contraction operator along the third modality, i.e., a matrix multiplication along the third modality of the \gls{hsi} tensor $\tens{X}$, and $\mathbf{A}\in\mathbb{R}^{L \times K}$ is the transmittance response of our acquisition system.


    The goal of the inversion protocol is to estimate the value of $\tens{X}$. We propose to derive an estimation $\hat{\tens{X}}$ of the spectrum $\tens{X}$ by setting up a Bayesian problem~\cite{Idie13:book} with spectro-spatial regularization.
    We specifically aim to minimize a cost function in the form:
    \begin{equation}
    	\hat{\tens{X}} = \argmin{\tens{X}}{
    		\frac{1}{2} \|\tens{Y} - \mathbb{A}(\tens{X})\|^2_2
    		+
    		\lambda g(\mathbb{L}\tens{X})
    	}\;,
    	\label{eq:regularized_techniques}
    \end{equation}
    where the right hand side is expressed as the sum of a data fidelity term and a term which characterizes the \textit{a priori} knowledge on the spectrum. In the above equation, $\|\cdot\|_2$ defines the $\ell_2$ norm applied over the concatenation of all pixels of its argument, while $g(\cdot)$, $\mathbb{L}$, and $\lambda>0$ denote a scalar functional, a linear operator and a regularization parameter, respectively.

    In the following sections we will detail the direct transformation and the regularization term.


\subsection{Interferometer transmittance function}

    The interferometric acquisition system is characterized by a given transmittance response $\matr{A}$.
    In the context of interferometry, which is the focus of our work, this defines a transformation from the spectral domain to an interferogram.
    For \glspl{fts}, it is common practice to express the spectral domain in terms of wavenumbers, i.e., the reciprocal of wavelengths.
    This spectrum is continuous in nature, but can be represented with a sufficiently fine sampled discretization of the wavenumber range $\{\sigma_k\}_{k\in[1,...,K]}$.
    On the contrary, the set of \glspl{opd} $\{\delta_l\}_{l\in[1,...,L]}$ where the interferogram is sampled is limited by the capability of the instrument of generating $L$ different optical paths. 
    While many techniques are available to measure an interferogram from a spectrum~\cite{picone2023interferometer, pisani-2009-compac-fourier, Hari03}, we only remind here that for Fabry-Perot interferometers the coefficients $a_{ik}$ of the transmittance response $\mathbf{A}$ can be modeled with an \textit{Airy distribution}~\cite{picone2023interferometer, Hech16}:

    \begin{equation}
        a_{lk} =
        \frac {(1-\mathcal{R})^2} {
            1 + \mathcal{R}^2
            -
            2\mathcal{R}
            \cos(2 \pi \sigma_k \delta_l)
        }\;,
        \label{eq:infty_wave}
    \end{equation}
    where $\mathcal{R}$ denotes the reflectivity of the surface of the interferometric cavities.
    

\subsection{Proposed reconstruction algorithm}
\label{ssec:proposed}

In this work, we propose to describe the linear operator $\mathbb{L}(\tens{X}) = \mathbb{L}^{(\mathrm{TV})}(\mathbb{L}^{(\mathrm{DCT})} (\tens{X}))$ from eq.~\eqref{eq:regularized_techniques} as the cascade of two operators $\mathbb{L}^{(\mathrm{DCT})}$ and $\mathbb{L}^{(\mathrm{TV})}$, which respectively operate in the spectral and spatial domain.
We define them as follows:

\begin{enumerate}
    \item \textit{Spectral operator $\mathbb{L}^{(\mathrm{DCT})}$}: we define it as type-II \gls{dct}. Specifically, we define a square matrix $\matr{W} \in \RR{K}{K}$, whose coefficients are:
    \begin{equation}
    		w_{ij} = \sqrt{\frac{2}{K}}
    		\cos\left(\frac{\pi}{K} \left(j-\frac{1}{2}\right) (i-1)\right)
    		_{\forall i, j \in \{1, \dots, K\}}\;.	
    	\label{eq:dct}
    \end{equation}
    The matrix is applied to the third modality of the latent \gls{hsi} $\tens{X}$ and we denote the result of this operation by $\tens{Z} = \mathbb{L}^{(\mathrm{DCT})}(\tens{X})=\matr{W} \con_3 \tens{X}$. The choice is taken as a follow-up of the results of our previous work~\cite{jouni2023model}, as the \gls{dct} has been shown to be particularly effective to induce sparsity on the spectrum.

    \item \textit{Spatial operator $\mathbb{L}^{(\mathrm{TV})}$}: we define it as \gls{tv} using the formulation of~\cite{Cond17}. The operator is applied on the spatial modalities of $\tens{Z}$ and returns a tensor $\tens{T} \in \RRRR{I}{J}{K}{2}$ such that $\forall \; i, \; j, \; k$:
    \begin{equation}
        \begin{cases}
        \begin{aligned}
            \tens{T}_{i,\; j,\; k,\; 1} = \tens{Z}_{i+1,\; j,\; k} - \tens{Z}_{i,\; j,\; k}\;,
            \\
            \tens{T}_{i,\; j,\; k,\; 2} = \tens{Z}_{i,\; j+1,\; k} - \tens{Z}_{i,\; j,\; k}\;.
        \end{aligned}
        \end{cases}
    \end{equation}
    In other words, this is the concatenation of the horizontal (columns) and the vertical (rows) differential operators, along a newly introduced fourth modality.
\end{enumerate}
%

For the regularization term $g(\cdot)$, we propose the collaborative norm $\ell_{1, 1, 1, 2}$, i.e., by imposing the $\ell_2$ norm on the fourth modality of $\tens{T}$ and the $\ell_1$ norm on the remaining three modalities.
This procedure, known as \gls{ctv}, was originally proposed in~\cite{DuraMSC16:jis} to combine different modalities, although their original formulation did not include any spectral operator.

The problem is then solved by primal-dual splitting algorithm known as Loris-Verhoeven~\cite{LoriV11:ip} with over-re\-laxation~\cite{CondKCH23:siam}, aiming to minimize the fidelity term of eq.~\eqref{eq:regularized_techniques}, where we plugged the \gls{dct} and \gls{tv} priors in a collaborative environment.
Algorithm \ref{algo:lv_updates} describes the iterative updates of our proposed method.
$\mathbb{A}^{\T}$ and $\mathbb{L}^{\T}$ represent the adjoint operators of $\mathbb{A}$ and $\mathbb{L}$, respectively, while $\|\mathbb{A}\|_{op}$ and $\|\mathbb{L}\|_{op}$ represent their operator norms~\cite{CondKCH23:siam}.
$\tens{X}$ and $\tens{U}$ are the primal and dual variables, respectively.
In our context, $\textrm{prox}_{\lambda, g^{\star}}(\tens{T})$ defines the proximal operator associated to the Fenchel conjugate of the $\ell_{1, 1, 1, 2}$ norm~\cite{PariB14:fto}.
$\forall \, i, \, j, \, k$, and denoting by $\tens{T}_{i, \; j, \; k, \; :}$ the $(i, j, k)$-th vector of $\tens{T}$ of dimension $2$ (along the fourth modality), this operator becomes:
	\begin{equation}
        \textrm{prox}_{\lambda, g^{\star}} (\tens{T}_{i, j, k, :}) =
        \begin{cases}
            \lambda\frac{\tens{T}_{i, j, k, :}}{\|\tens{T}_{i, j, k, :}\|_{2}} & \textrm{if } \|\tens{T}_{i, j, k, :}\|_2 > \lambda
            \\
            \tens{T}_{i, j, k, :} & \textrm{if } \|\tens{T}_{i, j, k, :}\|_2 \leq \lambda
        \end{cases}
        \label{eq:prox_g}
	\end{equation}

The values of $\eta$, $\tau$, and $\rho$ were chosen according to the relevant literature to ensure convergence \cite{CondKCH23:siam}.

\begin{algorithm}[t]
	\caption{
		Proposed method by \acrlong{lv} \cite{LoriV11:ip}
	}
	\begin{algorithmic}
		\REQUIRE $\mathbb{A}$, $\mathbb{L}$, $N_{\textrm{iters}}$
		
		\STATE
		\textbf{Initialize} \; $\tens{X}^{(0)} = \mathbb{A}^{\T} \tens{Y}$, \;$\tens{U}^{(0)} = \mathbb{L} \tens{X}^{(0)}$
		\STATE
		\textbf{Initialize} \; $\tau$=$0.99/\|\mathbb{A}\|^2_{op}$, $\eta$=$1/(\tau\|\mathbb{L}\|^2_{op})$, and $\rho$=$1.9$
		\STATE
		\textbf{Define} \; $\textrm{prox}_{\lambda\, g^{\star}} (\cdot)$ as in eq.~\eqref{eq:prox_g}
		\WHILE{\; Stopping criterion is not met \;}
		\STATE
		$
		\tens{V}^{(q)} = 
		\mathbb{A}^{\T} (
			\mathbb{A} \tens{X}^{(q)} - \tens{Y}
		)
		$
		\STATE
		$
		\tens{X}^{(q+\frac{1}{2})} = 
		\tens{X}^{(q)} - \tau \left(
			\tens{V}^{(q)}
			+
			\mathbb{L}^{\T} \tens{U}^{(q)}
		\right)
		$
		\STATE
		$
		\tens{U}^{(q+\frac{1}{2})} = 
		\textrm{prox}_{\lambda, g^{\star}} \left(
			\tens{U}^{(q)}
			+
			\eta \mathbb{L} \tens{X}^{(q+\frac{1}{2})}
		\right)
		$
		\STATE
		$
		\tens{X}^{(q+1)} = 
		\tens{X}^{(q)} - \rho\tau \left(
			\tens{V}^{(q)}
			+
			\mathbb{L}^{\T} \tens{U}^{(q+\frac{1}{2})}
		\right)
		$
		\STATE
		$
		\tens{U}^{(q+1)} = 
		\tens{U}^{(q)} + \rho \left(
			\tens{U}^{(q+\frac{1}{2})} - \tens{U}^{(q)}
		\right)
		$
		\ENDWHILE
		
		
		\RETURN $\hat{\tens{X}} = \tens{X}^{(N_{\textrm{iters}})}$  
	\end{algorithmic}
	\label{algo:lv_updates}
\end{algorithm}

\section{Experiments and results}
\label{sec:experiments_and_results}
\begin{figure}[t]
    \centering
    \begin{subfigure}[c]{0.45\linewidth}
        \centering
        \includegraphics{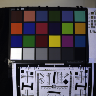}
        \caption{Reference}
        \label{fig:reference}
    \end{subfigure}
    \hfill
    \begin{subfigure}[c]{0.45\linewidth}
        \centering
        \includegraphics{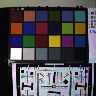}
        \caption{ADMM-TV}
        \label{fig:scico_tv}
    \end{subfigure}\\
    
    \begin{subfigure}[b]{0.45\linewidth}
        \centering
        \includegraphics{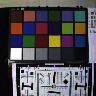}
        \caption{ADMM-BM3D}
        \label{fig:scico_bm3d}
    \end{subfigure}
    \hfill
    \begin{subfigure}[b]{0.45\linewidth}
        \centering
        \includegraphics{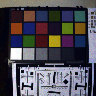}
        \caption{LV-DCT}
        \label{fig:lv_dct}
    \end{subfigure}\\
    
    \begin{subfigure}[b]{0.45\linewidth}
        \centering
        \includegraphics{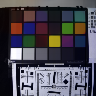}
        \caption{LV-TV}
        \label{fig:lv_tv}
    \end{subfigure}
    \hfill
    \begin{subfigure}[b]{0.45\linewidth}
        \centering
        \includegraphics{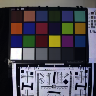}
        \caption{LV-CTV (ours)}
        \label{fig:lv_ctv}
    \end{subfigure}
    \caption{Reconstruction results with different inversion methods, using optimized regularization and convergence parameters. For visualization, three channels from the \gls{hsi} are chosen close to the central RGB wavelengths.}
    \label{fig:reconstruction}
\end{figure}

\begin{figure}[t]
    \centering
    \begin{subfigure}[b]{0.69\linewidth}
        \centering
        \includegraphics[width=\linewidth]{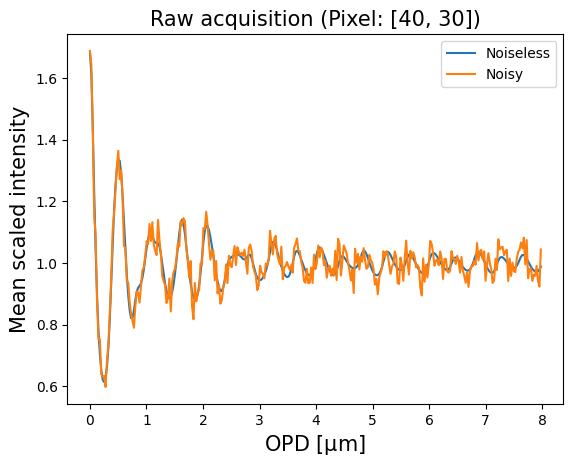}
        \caption{Acquisition}
        \label{fig:spectrum_acquisition}
    \end{subfigure}\\
    \begin{subfigure}[b]{0.69\linewidth}
        \centering \includegraphics[width=\linewidth]{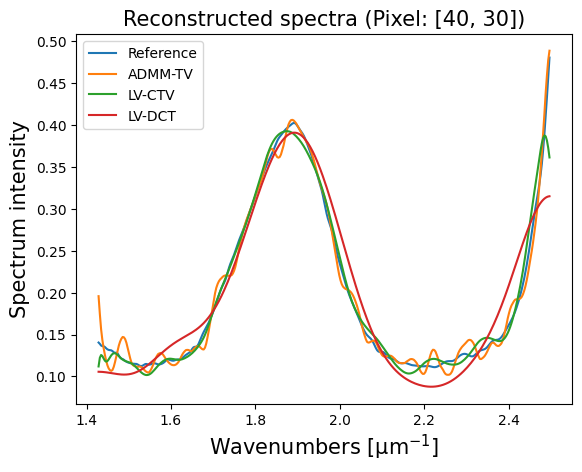}
        \caption{Reconstruction}
        \label{fig:spectrum_reconstruction}
    \end{subfigure}

    \caption{Visualization of the acquisition and the reconstruction for the central pixel of the green colorchecker box. (\subref{fig:spectrum_acquisition}) Comparison between noiseless and noisy acquisition with \glsentryshort{snr} of $10 \mathrm{dB}$. (\subref{fig:spectrum_reconstruction}) Comparison of reconstructed spectra with a selection of the inversion methods described in this paper.}
    \label{fig:spectrum}
\end{figure}
\subsection{Experimental setup}

    To test the effectiveness of the reconstruction algorithms, we operate in a simulated scenario. 
    Specifically, we performed our tests using a \gls{hsi} that we acquired with the commercial hyperspectral camera Specim IQ, whose scene contains a colorchecker and some geometrical targets.
    To generate our reference, we spatially resized the raw acquisition to $96 \times 96$ pixels and interpolated the channels to regularly sample the range of visible wavenumbers. 
    This range is between $[1.4, 2.5]$ \si{\um}$^{-1}$ and contains $366$ channels.
    We then generated a transmittance response for our interferometric camera using eq.~\eqref{eq:infty_wave}, using parameters replicating the setup of Pisani and Zucco in~\cite{pisani-2009-compac-fourier}. We made this choice as their setup is already optimized for reconstructing spectra in the visible range and we assume for simplicity that the the interferogram can be sampled even at very low \glspl{opd}. 
    More specifically we chose the interferogram domain to be regularly sampled within an \gls{opd} range in $[0,\; 8]$ \si{\um} with $25$ \si{\nm} step, so that each pixel is represented by 319 samples; the reflectivity is set to $\mathcal{R}=0.255$.
    Finally, we add Gaussian noise to the simulated acquisition obtaining a tensor $\tens{Y}\in\mathbb{R}^{96 \times 96 \times 319}$, so that the mean-centered version of the obtained signal has a \gls{snr} of 10 \si{\decibel}.
    This signal is processed with a series of techniques that will be discussed in the next section, in order to reconstruct an image datacube $\hat{\tens{X}}\in\mathbb{R}^{96 \times 96 \times 366}$, to be compared with our same-sized reference $\tens{X}$. The objective analysis will be performed using the \gls{ssim} and \gls{rmse} quality indices.



\subsection{Results and discussion}

    \begin{table}[t]
    \centering
    \caption{Objective quality assessment reconstruction results.}
    \begin{tabular}{l|c|c}
        Method & RMSE [$\times 10^{-2}$] & SSIM \\\hline
        Reference & 0 & 1\\\hline
        \glsentryshort{admm}-\glsentryshort{tv} & 3.101 & 0.9857 \\
        \glsentryshort{admm}-\glsentryshort{bm3d}~\cite{Venk13} & 3.208 & 0.9853 \\
        \gls{lv}-\gls{dct} \cite{jouni2023model} & 3.046 & 0.9716 \\
        \gls{lv}-\gls{tv} & 2.737 & 0.9794 \\
        \gls{lv}-\gls{ctv} (ours) & \textbf{1.523} & \textbf{0.9946} \\
    \end{tabular}
    \label{tab:reconstruction}
\end{table}

    The reconstruction algorithms used in our tests employ two frameworks. 
    The first one employs our own implementation of the \glsreset{lv}\gls{lv} algorithm~\cite{LoriV11:ip}, in the over-relaxed version~\cite{CondKCH23:siam} that we used in a previous work~\cite{jouni2023model}.
    In this framework we tested three different linear operators for the regularizer $\mathbb{L}$ to prove the effectiveness of the proposed inversion strategy described in Section~\ref{ssec:proposed}. In particular, we tested our setup both using the each of the \gls{dct} operator $\mathbb{L}^{(\mathrm{DCT})}$ and the \gls{tv} operator $\mathbb{L}^{(\mathrm{TV})}$ separately, then using them in cascade as described in Section~\ref{ssec:proposed}. The three methods are denoted respectively as \gls{dct}, \gls{tv} and \gls{ctv} in figures and tables.
    We also tested our reconstruction problem employing a \gls{pnp} strategy~\cite{Venk13}. For this purpose, eq.~\eqref{eq:regularized_techniques} was solved with the linearized \gls{admm}~\cite{Yang12} implementation in SCICO~\cite{Balk22}, where the proximal operator associated to the reconstruction is substituted with a general purpose denoiser. We tested this strategy with both an isotropic \gls{tv} and a \gls{bm3d}~\cite{Dabo08} denoiser.
    For all tested reconstruction methods the hyperparameters were adjusted to the best of our abilities and the tests were ran for 100 iterations, except for the \gls{bm3d}, which was run for 10 iterations as it is much slower than the counterparts.
    The objective quality results, shown in \tablename~\ref{tab:reconstruction}, show the benefit of the joint spectral and spatial regularization, especially in the \gls{lv} framework.

    The results are confirmed by the visual comparison and given in \figurename~\ref{fig:reconstruction} and~\ref{fig:spectrum}. The application of \gls{tv} allows to strongly improve the spatial coherence of the reconstructed image. 
    This is an expected result for the image that we used for our test, which is mostly patch based, for which the \gls{tv} is known to have competitive performances.
    On the contrary, strong distortions are still present with \gls{pnp} based algorithms. We should however note that we are operating in a disadvantageous scenario for them to be applied in their unaltered form, as \gls{bm3d} is typically better suited for images with higher spatial resolution, that enhance its capability to derive common spatial features on the scene. 
    This shows the importance of using priors for the regularizers that are adapted to the data.
    In general terms, the \gls{lv} framework converges faster than the \gls{admm}, as proven by the results with the \gls{tv} regularizer.






\section{Conclusion}

    In this work, we have shown the importance of the spatial regularization for the reconstruction of datacubes acquired in the domain of the interferograms. In future works, we plan to extend these results by further investigating \gls{pnp} methods adapted for hyperspectral data, exploring non-redundant domain transformations, such as tensor decompositions.

\vfill
\clearpage

\bibliographystyle{IEEEbib}
\bibliography{strings,refs}

\end{document}